\documentclass{iau} 
\usepackage{graphicx,url}

\title[MUSE observations of 1E0102] 
{MUSE Integral Field Observations of the Oxygen-rich SNR 1E\,0102.2-7219}

\author[Seitenzahl et al.]   
{Ivo R. Seitenzahl$^{1,2,3}$
  \and Fr\'ed\'eric P.~A.~Vogt$^4$ \and Jason P.~Terry$^5$ \and Michael A. Dopita$^2$ \and Ashley J.~Ruiter$^{2,3}$ \and Parviz Ghavamian$^6$ \and Tuguldur Sukhbold$^7$}

\affiliation{$^1$ School of Physical, Environmental and Mathematical Sciences, University of New South Wales, Australian Defence Force Academy, Canberra, ACT 2600, Australia.\\       email: {\tt i.seitenzahl@adfa.edu.au} \\ [\affilskip]
 $^2$ Research School of Astronomy and Astrophysics, Australian National University, Canberra, Australia.\\ [\affilskip]
$^3$ ARC Centre for All-sky Astrophysics (CAASTRO). \\ [\affilskip]
  $^4$ European Southern Observatory, Av. Alonso de C\'ordova 3107, 763 0355 Vitacura, Santiago, Chile. \\ [\affilskip]
  $^5$ Department of Physics and Astronomy, University of Georgia, USA.\\ [\affilskip]
  $^6$ Department of Physics, Astronomy and Geosciences, Towson University, Towson, MD 21252, USA.\\ [\affilskip]
$^7$ Department of Astronomy  and Center for Cosmology \&
Astro-Particle Physics, The Ohio State University, Columbus,
OH 43210, USA.\\}
\pubyear{2017}
\volume{331}  
\jname{SN 1987A, 30 years later -- Cosmic Rays and Nuclei from Supernovae and their aftermaths}
\editors{M.~Renaud, A.~Marcowith, G.~Dubner, A.~Ray \& A.~Bykov, eds.}
\begin{document}

\maketitle

\begin{abstract}
We have observed the oxygen-rich SNR 1E\,0102.2-7219 with the integral field spectrograph WiFeS at Siding Spring Observatory and discovered sulfur-rich ejecta for the first time. Follow-up deep DDT observations with MUSE on the VLT (8100\,s on source) reaching  down to a noise level of ${\sim}5 \times 10^{-20} \mathrm{erg}\mathrm{s}^{-1}\mathrm{cm}^{-2}\mathrm{\AA}^{-1}\mathrm{spaxel}^{-1}$ have led to the additional discovery of fast-moving hydrogen as well as argon-rich and chlorine-rich material. The detection of fast-moving hydrogen knots challenges the interpretation that the progenitor of 1E\,0102 was a compact core of a Wolf-Rayet star that had shed its \textit{entire} envelope. In addition to the detection of hydrogen and the products of oxygen-burning, this unprecedented sharp (0.2$^{\prime\prime}$ spaxel size at ${\sim}0.7^{\prime\prime}$ seeing) and deep MUSE view of an oxygen-rich SNR in the Magellanic Clouds reveals further exciting discoveries, including [Fe\,\textsc{xiv}]$\lambda$5303 and [Fe\,\textsc{xi}]$\lambda$7892 emission, which we associate with the forward shock. We present this exciting data set and discuss some of its implications for the explosion mechanism and nucleosynthesis of the associated supernova.

\end{abstract}

\firstsection 
\section{Introduction}
\textbf{1E 0102.2-7219} (1E\,0102 for short) belongs to the class of \textit{oxygen-rich} young supernova remnants (O-rich YSNRs) showing remarkably little-to-no hydrogen emission. The prominent oxygen forbidden-line emission observed in the filaments and knots (see Figs.~1,~2) is understood to arise from shocked ejecta from a core-collapse supernova that had been stripped of (most of) its hydrogen envelope prior to explosion. Unlike most SNRs, where the optical emission comes from interstellar medium ionized by the forward shock, O-rich YSNRs are special in that one can study the actual ejecta composition of the supernova, providing unique and direct insights into explosion mechanisms and nucleosynthesis conditions in core-collapse supernovae. Only a handful of such O-rich YSNRs are known: the best studied examples include Cas A in our Galaxy and 1E\,0102 in the Small Magellanic Cloud (SMC).
\clearpage
Since the discovery of the high-velocity oxygen-rich supernova ejecta in 1E\,0102 (\cite{dopita1981a}), subsequent searches (including HST FOS spectroscopy and WFC3 imaging) for the expected products of oxygen burning, in particular sulfur, have returned negative results (e.g., \cite{lasker1991a,blair2000a}).  These searches have only detected carbon, oxygen, neon, and magnesium, but no products of oxygen-burning, i.e., no sulfur, silicon, calcium, or argon. The angular size of 1E\,0102 (50$^{\prime\prime}$ in diameter) and the large Doppler shifts associated with its fast-moving ejecta ($v \gtrsim 1000\,\mathrm{km\,s}^{-1}$) have been strong limitations for previous long-slit spectroscopic and narrow-band imaging searches, which have either failed to map locations of interest and/or were blind to fast-moving ejecta altogether. 

\section{WiFeS observations}
\begin{figure}[h]
\begin{center}
 \includegraphics[width=\columnwidth,trim= 0 0 80 40,clip]{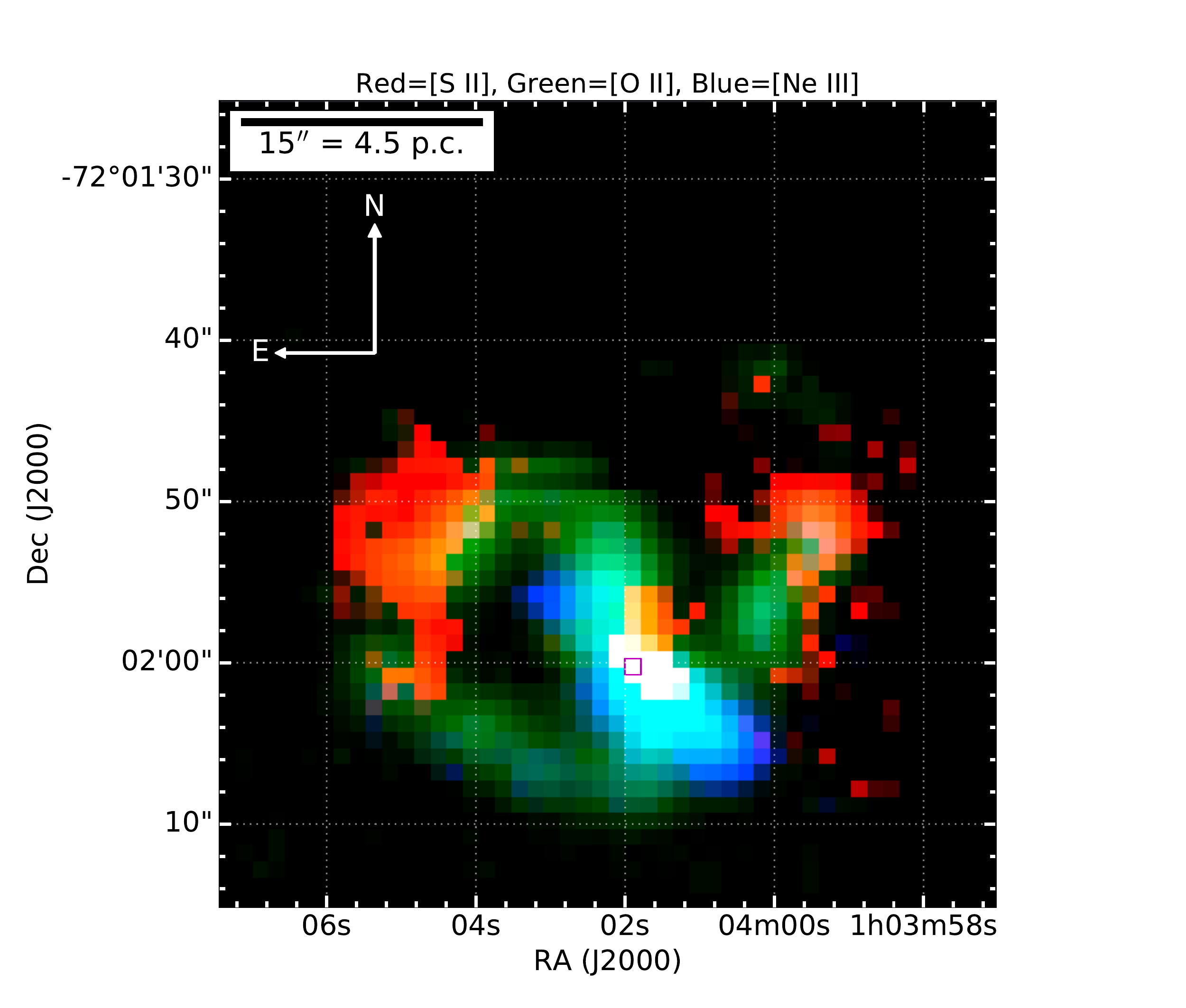} 
 \caption{Three color (RGB) image of 1E\,0102.2-7219 showing the regions where fast-moving [S\,\textsc{ii}]$\lambda\lambda$6716,6731 (red), [O\,\textsc{ii}]$\lambda\lambda$3726,3729 (green), and [Ne\,\textsc{iii}]$\lambda\lambda$3869,3967 (blue) ejecta is detected with WiFeS.}
   \label{fig1}
\end{center}
\end{figure}

In our observations (P.I.: Seitenzahl) on the nights of Aug.~7 and Aug.~8, 2016 with the WiFeS Integral Field Spectrograph on the 2.3\,m telescope at Siding Springs Observatory, we have detected fast-moving (mostly, but not exclusively blue-shifted) [S\,\textsc{ii}]$\lambda\lambda6716,6731$ emission (see Fig.\,\ref{fig1}). These data represent the first detection in the optical of the products of oxygen-burning in the ejecta of 1E\,0102. The first set of observations were performed with the R7000 and B7000 gratings in place and constituted a co-added $2 \times 1800\,\mathrm{s}$ science observation block for each of the two pointings that make up the mosaic, with a 900\,s dark sky frame before and after. We re-visited the same two fields of the target for an additional $2\times1800\,\mathrm{s}$ per science field on the nights of Oct.~28 and Oct.~29, 2016, this time with the B3000 in the blue arm to include [O\,\textsc{ii}]$\lambda\lambda$3726,3729 and [Ne\,\textsc{iii}]$\lambda\lambda$3869,3967 (see Fig.~2). All data were reduced and flux-calibrated with the PyWiFeS pipeline (\cite{childress2014a}).
\begin{figure}[h]
\begin{center}
 \includegraphics[width=\columnwidth, trim= 30 0 100 0,clip]{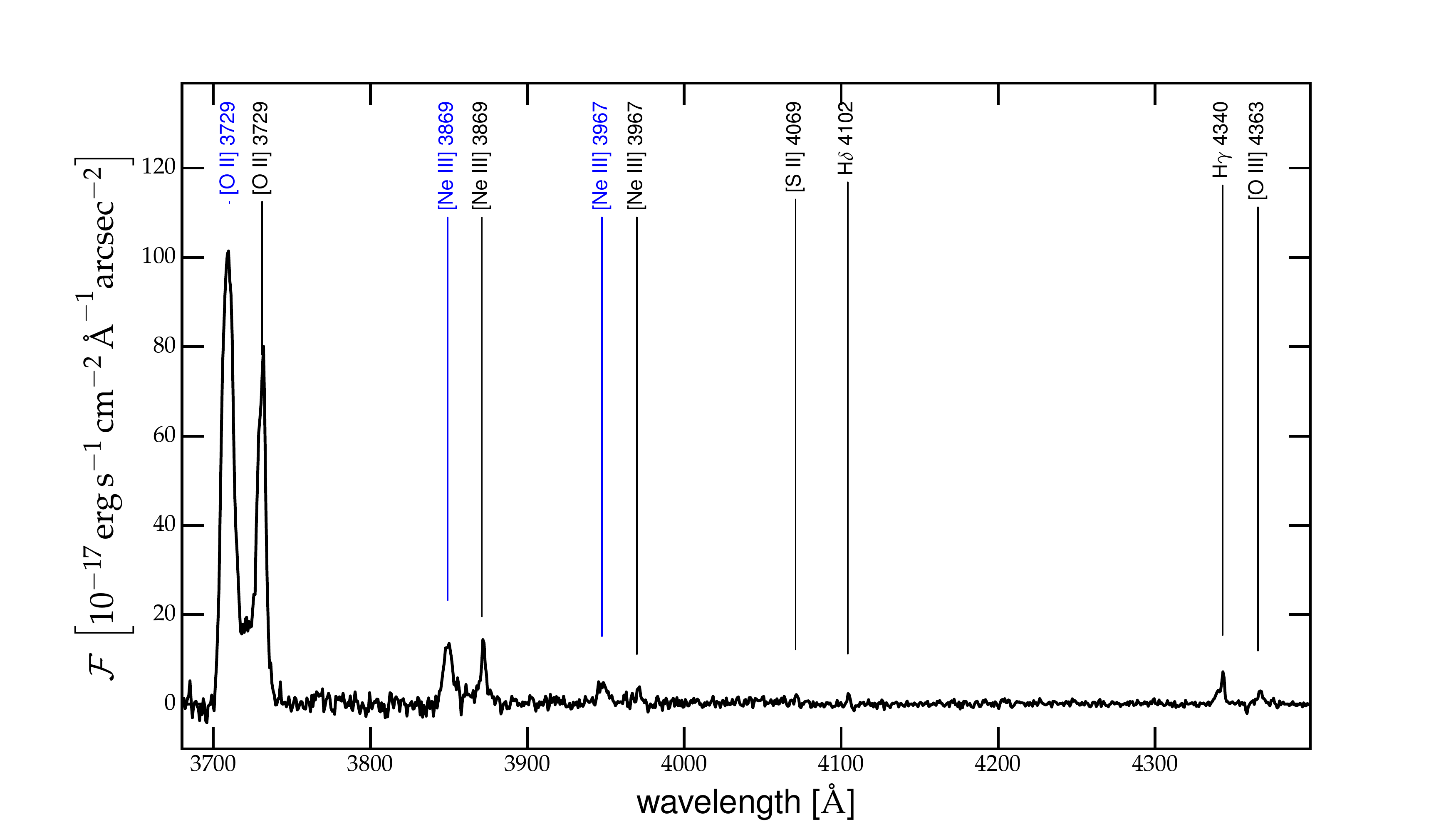} 
 \caption{Blue end of the WiFeS spectrum of the bright knot indicated by the purple square in Fig.~1. The broad, fast-moving ejecta component is labeled in blue ([O\,\textsc{ii}] and [Ne\,\textsc{iii}] inaccessible with MUSE), the narrow line emission at the local velocity of 1E\,0102 is labeled in black.}
   \label{fig2}
\end{center}
\end{figure}

\section{MUSE observations}
\begin{figure}[h!]
\begin{center}
 \includegraphics[width=\columnwidth]{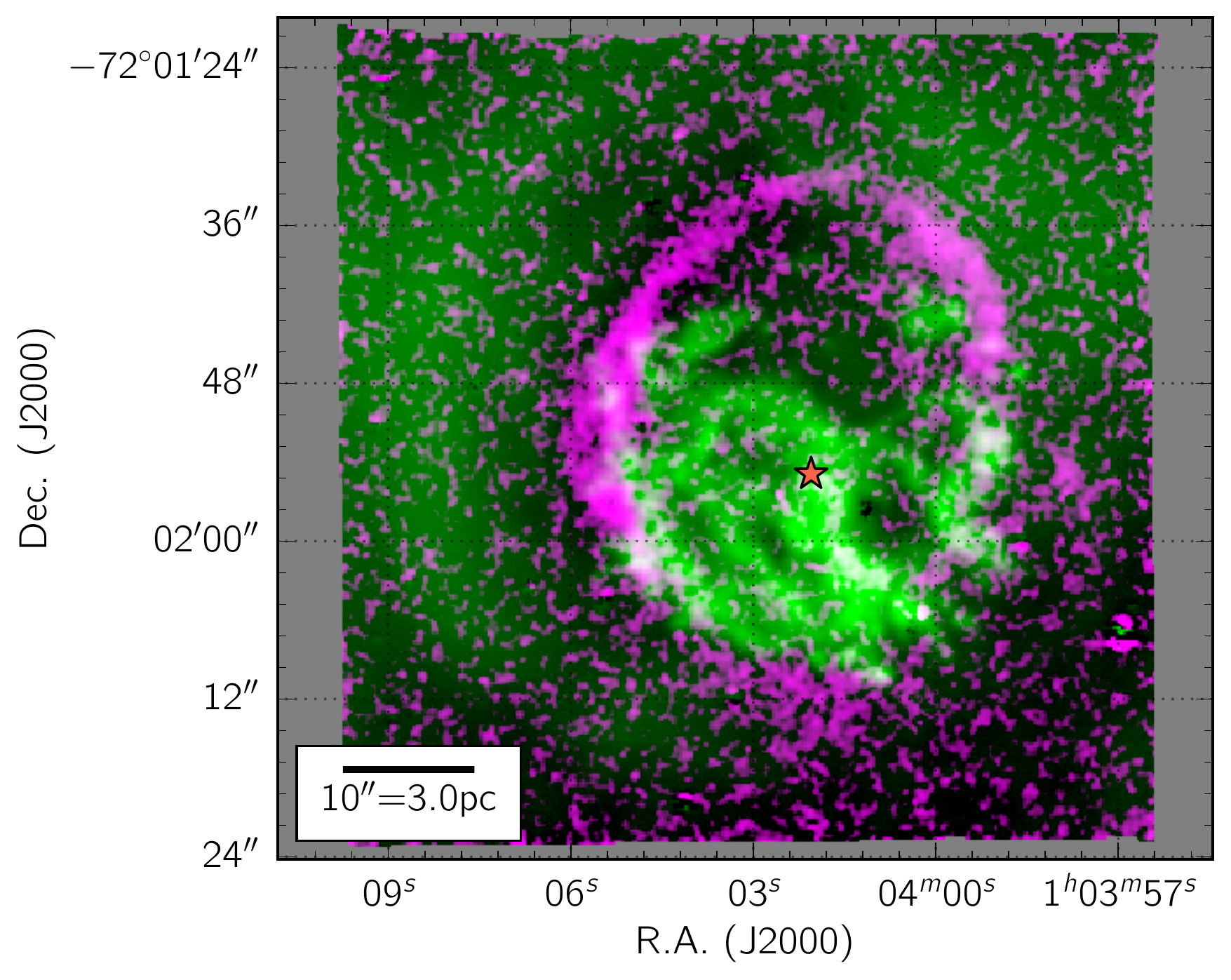} 
 \caption{MUSE view of 1E\,0102, reproduced from \cite{vogt2017b}. Shown in magenta is [Fe\,\textsc{xiv}] (5300 -- 5308 \AA\ slice through the cube) and in green [O\,\textsc{iii}] (4900 -- 5100 \AA\ slice through the cube). The center of expansion derived by \cite[Finkelstein et al.\ (2006)]{finkelstein2006a} is marked with a star symbol. The clumpy O-rich ejecta is seen in bright green mostly inside the magenta [Fe\,\textsc{xiv}] ring. The more diffuse [O\,\textsc{iii}] emission outside the magenta [Fe\,\textsc{xiv}] ring is attributed to the photoionization precursor ahead of the forward shock.}
   \label{fig3}
\end{center}
\end{figure}

We observed 1E 0102 with the Multi-Unit Spectroscopic Explorer (MUSE) in Service Mode at the Very Large Telescope on the night of October 7, 2016 via Director's Discretionary Time (DDT) program 297.D-5058 (P.I.: F.P.A. Vogt). We obtained a total of $9 \times 900\,\mathrm{s}$ exposures on-source, with appropriate 180\,s dark sky exposures away from the SMC in between the science exposures. The data frames were individually reduced and then combined into a single data cube with the ESO Reflex workflow (\cite{freudling2013a}), for further details see \cite{vogt2017b}. Compared to the 1$^{\prime\prime}$ by 1$^{\prime\prime}$ spaxel size and ${\sim}$1.5$^{\prime\prime}$ seeing of the WiFeS cube, the 0.2$^{\prime\prime}$ by 0.2$^{\prime\prime}$ spaxel size and ${\sim}$0.7$^{\prime\prime}$ seeing of the MUSE cube offers a much sharper view of 1E\,0102 (see Fig.~3). In addition to gaining in spatial resolution, the MUSE observations have increased the sensitivity by a factor of approximately 100, improving from the WiFeS noise level of a few times $10^{-18}\,\mathrm{erg}\mathrm{s}^{-1}\mathrm{cm}^{-2}\mathrm{\AA}^{-1}\mathrm{spaxel}^{-1}$ to a noise level of a few times $10^{-20}\,\mathrm{erg}\mathrm{s}^{-1}\mathrm{cm}^{-2}\mathrm{\AA}^{-1}\mathrm{spaxel}^{-1}$.
\begin{figure}[h!]
\begin{center}
 \includegraphics[width=\columnwidth,trim=125 0 135 0, clip]{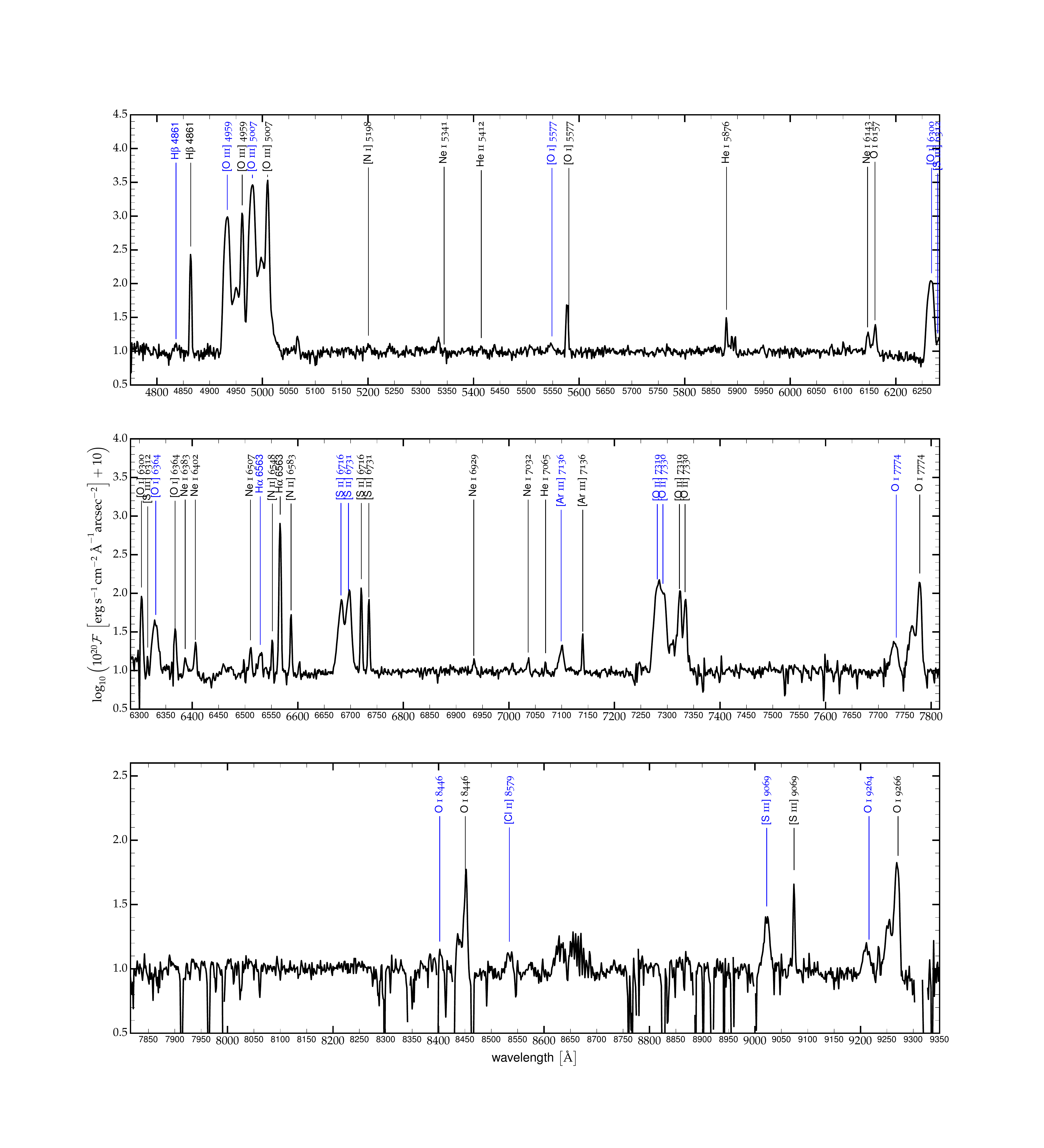} 
 \caption{MUSE spectrum of the same S-bright knot as for the WiFeS spectrum in Figs.~1 and 2. The broad, fast-moving ejecta component is labeled in blue, the narrow line emission at the local velocity of 1E\,0102 is labeled in black. Note in particular the clear detections of blue-shifted [S\,\textsc{ii}]$\lambda\lambda$6716,6731, [S\,\textsc{iii}]$\lambda$9069, [Ar\,\textsc{iii}]$\lambda$7136, and the (marginal) detection of [Cl\,\textsc{ii}]$\lambda$8579. Weak lines of H$\alpha$ and H$\beta$ are seen at the same Doppler shift. }
   \label{fig4}
\end{center}
\end{figure}

\section{Summary}
In this contribution, we have presented optical integral field spectroscopy observations of 1E\,0102.2-7219 performed with the WiFeS and MUSE instruments. The combined data cubes extend from $3500\,\mathrm{\AA}$ to $9350\,\mathrm{\AA}$. In addition to mapping out the (previously known) O- and Ne-rich ejecta in unprecedented detail, we have discovered fast-moving [S\,\textsc{ii}], [S\,\textsc{iii}], [Ar\,\textsc{iii}], H$\alpha$, H$\beta$, and possibly [Cl\,\textsc{ii}], which we attribute to be components of the supernova ejecta. S, Ar, and Cl are main nucleosynthesis products of O-burning, which we detect here as ejecta in the optical in 1E\,0102 for the first time. The discovery of Balmer line emission in the ejecta indicates that the progenitor star must not have been entirely stripped of its hydrogen envelope when it exploded. We have also discovered high ionization coronal lines of Fe ([Fe\,\textsc{x}], [Fe\,\textsc{xi}], and [Fe\,\textsc{xiv}]) and the photoionization precursor emission, which can be used to derive the forward shock parameters (see \cite{vogt2017b}).

\clearpage

\section{Acknowledgements}

This research has made use of \textsc{brutus}, a Python module to process data cubes from integral field spectrographs hosted at \url{http://fpavogt.github.io/brutus/}. For this analysis, \textsc{brutus} relied on \textsc{statsmodel} (\cite{Seabold2010}), \textsc{matplotlib} (\cite{Hunter2007}), \textsc{astropy}, a community-developed core Python package for Astronomy (\cite{AstropyCollaboration2013}), \textsc{aplpy}, an open-source plotting package for Python hosted at \url{http://aplpy.github.com}, and \textsc{montage}, funded by the National Science Foundation under Grant Number ACI-1440620 and previously funded by the National Aeronautics and Space Administration's Earth Science Technology Office, Computation Technologies Project, under Cooperative Agreement Number NCC5-626 between NASA and the California Institute of Technology. This research has also made use of the \textsc{aladin} interactive sky atlas (\cite{Bonnarel2000}), of \textsc{saoimage ds9} (\cite{Joye2003}) developed by Smithsonian Astrophysical Observatory, of NASA's Astrophysics Data System, and of the NASA/IPAC Extragalactic Database (NED; \cite{Helou1991}), which is operated by the Jet Propulsion Laboratory, California Institute of Technology, under contract with the National Aeronautics and Space Administration. IRS was supported in part by the Australian Research Council Laureate Grant FL0992131 and the Future Fellowship Grant FT160100028. AJR acknowledges funding through the ARC Centre of Excellence for All-sky Astrophysics (CAASTRO) through project number CE110001020. PG thanks the Stromlo Distinguished Visitor Programme. FPAV and IRS thank the CAASTRO AI travel grant for generous support. Based in part on observations made with ESO Telescopes at the La Silla Paranal Observatory under programme ID 297.D-5058[A] and the ANU 2.3\,m telescope at Siding Spring Observatory.

\begin{discussion}
\discuss{Kirshner}{It's great you finally found [S\,\textsc{ii}] -- in a way we always knew it had to be there. You should try to put a limit on [Ca\,\textsc{ii}].}
\end{discussion}

\end{document}